# A simple physical model for scaling in protein-protein interaction networks

Eric J. Deeds*, Orr Ashenberg† and Eugene I. Shakhnovich‡

*Department of Molecular and Cellular Biology, Harvard University, 7 Divinity Avenue, Cambridge, MA 02138, USA

†Harvard College, 12 Oxford Street, Cambridge, MA 02138, USA

‡Department of Chemistry and Chemical Biology, Harvard University, 12 Oxford Street, Cambridge, MA 02138, USA

# Summary

**Background**

It has recently been demonstrated that many biological networks exhibit a "scale-free" topology where the probability of observing a node with a certain number of edges ($k$) follows a power law: i.e. $p(k) \sim k^{-\gamma}$. This observation has been reproduced by evolutionary models.

**Principal Findings**

Here we consider the network of protein-protein interactions and demonstrate that two published independent measurements of these interactions produce graphs that are only weakly correlated with one another despite their strikingly similar topology. We then propose a physical model based on the fundamental principle that (de)solvation is a major physical factor in protein-protein interactions. This model reproduces not only the scale-free nature of such graphs but also a number of higher-order correlations in these networks. A key support of the model is provided by the discovery of a significant correlation between number of interactions made by a protein and the fraction of hydrophobic residues on its surface.

**Significance**

The model presented in this paper represents the first physical model for experimentally determined protein-protein interactions that comprehensively reproduces the topological features of interaction networks. These results have profound implications for understanding not only protein-protein interactions but also other types of scale-free networks.

## Introduction

Many studies in recent years have revealed that a large variety of systems, from the World Wide Web to the network of chemical reactions catalyzed in a cell, exhibit a particularly interesting "scale-free" topology when represented as graphs [1-6]. In these systems the probability of finding an object (or node) that connects $k$ other nodes in the graph follows a power-law; i.e. the degree distribution (or $p(k)$) has the form $p(k) \sim k^{-\gamma}$ [1]. This observation has (in general) been explained in terms of dynamical models based on the principles of network growth and an effective "preferential attachment" whereby objects that have many links at some point in time are more likely to acquire nodes as the graph grows than objects with fewer connections [1,7]. The fact that scale-free networks are so often observed in biological systems has lead to the proposal that many evolutionary processes exhibit mechanisms similar to preferential attachment that are based on the duplication and divergence of genes [4,8-11].

One of the biological networks that has undergone considerable study is the set of interactions between proteins in the cell. The advent of high-throughput methods for measuring the binding of one protein to another using the yeast two-hybrid (Y2H) system has allowed for the characterization of large numbers of interactions between proteins in organisms such as *Saccharomyces cerivisiae*, *Helicobacter pylori*, *Caenorhabditis elegans* and *Drosophila melanogaster* [12-16]. Two major independent Y2H experiments have been performed to determine the "interactome" of *S. cerivisiae* [12,13], and graphs of these interactions reveal that these systems constitute scale free networks with power-law exponents ranging from ~2.0 to ~2.7 [1,3,12,13,17,18].

It has long been noted, however, that yeast two-hybrid screens are rather inaccurate and can lead to relatively "noisy" sets of interactions [19-22]. Indeed, when the two major *S. cerivisiae* protein-protein interaction experiments are compared with one another, one finds that only about 150 out of the 1000s of interactions identified in each experiment are recovered in the other experiment [22]. A similar lack of agreement has recently been found for independent Y2H experiments in D. melanogaster [23]. Although computational methods have been proposed that may allow for some reduction of noise, it is clear that the rate of both false positives and false negatives in these experiments may be quite high [19-22]. Moreover, it is known that when a protein is used as "bait" (i.e. fused to the DNA binding component of the yeast two-hybrid) it will tend to exhibit more interactions than when employed as "prey" [19]. It is thus very clear that these experiments may contain a large number of artifacts.

In the present work we have explored these potential artifacts by considering the hypothesis that the interactions reported by the Y2H method are dominated by non-specific interactions between proteins. This hypothesis is primarily motivated by our observation that, in general, the connectivity of a given protein is not well correlated between the Uetz and Ito experiments (see Figure 1). We propose an entirely physical model in order to explain how two networks with essentially uncorrelated connectivities could nonetheless display profoundly similar (scale-free) topologies. We demonstrate that this model, when combined with an elemental source of experimental noise, reproduces the degree distributions of the experimentally determined PPI networks. The exposure of random surfaces between experiments (and thus a varying number of hydrophobic residues that thermodynamically drive interactions) is sufficient to explain the lack of correlation between two experiments that exhibit scaling in their degree distributions. We further show that "higher-order" features of these networks, such as the scaling of the clustering coefficient of a node with its connectivity (i.e. $C$ as a function of $k$), are also recovered in this model. These results indicate that the observation of such topological features is not contingent on any specific evolutionary dynamics or evolutionary pressure for such networks to be "robust," "hierarchical" or "modular," as has been previously proposed. Finally we observe a strong correlation between the hydrophobicity of a protein and its number of interacting partners, a finding that is in complete agreement with our physical model. Together these results demonstrate that the protein-protein interactions as assayed by the Y2H techniques need not report only evolved and specific interactions, and that the interesting (non-random) topological features of these graphs need not have an evolutionary origin. Although our results do not indicate that these networks contain no evolutionarily or biologically important information, they do imply that a large number of observations in these and (perhaps) other biological networks might contain considerable influences from non-specific interactions.

## Results

### Correlations in the Number of Interacting Partners

To further explore the scale-free graphs obtained from these potentially noisy experiments we considered the graph of interactions between proteins for the 676 proteins that exhibited interactions in both the Uetz *et al.* and the Ito *et al.* experiments [12,13]. We then compared the number of interactions measured for a given protein in one of the assays to the number of interactions for that same protein observed in the other assay. As evidenced by Fig. 1A, the correlation between the degree of a given protein in the two experiments is quite weak with an $R^2$ of 0.18 for nodes of all degrees and an $R^2$ of 0.068 if the three outliers are ignored (i.e. considering only nodes of degree less than 20). The situation is much the same when the comparison is made using the more reliable ItoCore dataset [13] (Fig. 1B). These very low $R^2$ values are striking, considering that these represent the same proteins from the same organism assayed in very similar experiments, and it is clear that these two graphs, while topologically similar, are statistically unrelated. If one set of interactions is assumed to represent the "true" set of evolved protein-protein interactions in yeast, it follows that the other graph must consist largely of experimental noise, a finding that casts doubt on the reliability of either data set. Indeed, this observation may indicate why the number of interactions made by a protein in protein-protein interaction (PPI) networks is only very weakly correlated with evolutionary rates [24].

The fact that these networks are scale-free, however, rules out the possibility that apparent protein-protein associations in either case are entirely random: if they were, the graph would represent a random graph and one would observe a Poisson or Gaussian degree distribution in the resulting networks [1]. In order to reconcile these two observations, we posit a simple physical model of protein-protein interactions. First, we assume that much of the free energy of binding that characterizes a particular protein-protein interaction is due to the burial and desolvation of hydrophobic groups at the binding interface [25-27]. In this case, we hypothesize that the low correlation in connectivity between the two datasets is largely due to the exposure of different surfaces for each protein in each of the Y2H experiments.

**The MpK Model**

Suppose there are $N$ surface residues for a particular protein, and a given fraction $p$ of them are hydrophobic. Say that $M$ of those residues are actually exposed and involved in binding the other proteins in the experiment, and that $K$ out of those $M$ residues are hydrophobic. If we assume that $M$ is sampled from $N$ randomly and independently, it is clear that the probability of finding $K$ hydrophobic residues within $M$ follows a binomial distribution:

$$p(K) = \binom{M}{K} p^K (1-p)^{M-K}. \tag{1}$$

In this case, each protein-protein interaction will result in the burial of a certain total number of hydrophobic groups; i.e. $K_{ij} = K_i + K_j$ (see Fig. 2). The desolvation of $K_{ij}$ hydrophobic residues is related to the free energy of protein binding and represents a standard way to treat the strength of hydrophobic interactions [25-27]. In this case we simply take the free energy of binding $F_{ij}$ to be equal to $-K_{ij}$. The Y2H experiments are based on binding affinity, not binding free energy, and it follows from statistical mechanics and thermodynamics that the affinity $A_{ij}$ between two proteins $i$ and $j$ will follow $A_{ij} \sim \exp(-F_{ij})$ if we set the temperature scale of our experiment such that $kT = 1$. To build a PPI network we define an experimental limit of sensitivity $A_C$ corresponding to the weakest interaction (the interaction which buries the fewest hydrophobic groups) that is nonetheless sufficiently strong to be detected by the experiment.

In order to simulate this model we must first understand the distribution of $p$ values for proteins in the experiment. To do this we employ a simple homology modeling procedure (described in the Methods section of the Supporting Information) to transfer solvent accessibilities from proteins of known structures to their corresponding homologs from the Ito Y2H dataset. We find that this distribution is well fit by a Gaussian function (see Fig. 2B). In our model of the Y2H experiment we sample 3200 values of $p$ from a Gaussian distribution with the same mean and standard deviation (Fig. 2B). We use the same value of $M$ for each protein in the experiment given that the stereotypical size of the binding surface is not determined by the surface area of the protein itself but rather the average size of the interface across all the other proteins in the experiment. The choice of $M$ is essentially arbitrary (see the discussion of $A_C$ below and in the Supporting Information), and in the case of our results it is set to be 100.

We find that, within certain ranges of $A_C$, the networks created by this MpK model discussed above exhibit degree distributions that are well-fit by power-law functions; a representative example is shown in Fig. 2C (a discussion of the variance in the degree distributions for different realizations of this model may be found in the Supporting Information). This model indicates that, at stringent cutoffs, many of the nodes in the graph are orphans, a finding that fits well with experimental observations from both Uetz and the ItoCore data set [12,13] (note that, in contrast to the graphs from Fig. 3, orphans are displayed on the log-log plot in Fig. 2C by adding 1 to the degree of each node). This finding indicates that the apparent scaling in these systems could very easily arise from a set of completely non-specific interactions that contain no evolutionary information. $A_C$ determines the apparent power-law exponent $\gamma$ and is the only truly fitable parameter in the model; for any value of $M$ that is sufficiently large to capture the differences in $p$ that exist in the population, one may obtain a degree distribution of a given $\gamma$ simply by changing the value of $A_C$ (for further discussion of these points see the Supporting Information). Although this model is mathematically related to other static models of scale-free networks [28] it is important to note that our model represents the first model of PPI networks that attempts to consider the physics of protein binding and is based on a Gaussian distribution of some underlying property.

**Random Noise**

The above model, while suggestive, is not necessarily a complete model of all of the PPI experiments—for instance, in the case of the original Ito data set, the number of orphans is much smaller (the experiment reports many more connected nodes than our model predicts) and the degree distribution deviates from power-law behavior at small values of $k$ [6]. To better model both of these experimental observations we add an elemental source of noise to our model by linking a number of orphans to randomly chosen nodes in the graph. To model a particular dataset, we first fit the value of $A_C$ to the power-law exponent that is observed in the experimental data. We randomly connect a number of orphans to nodes in the graph in order to obtain a number of connected nodes in the graph exactly equal to the number of connected nodes in the dataset. The degree distributions of these random linking graphs exhibit surprisingly good agreement with the experimental results in all cases (see Fig. 3A for the Ito model and the Supporting Information for the ItoCore and Uetz models). The relationship between Ito and ItoCore is quite natural in this model—ItoCore is a dataset that is obtained at higher stringency (representing the greater number of colonies needed to count an interaction in the ItoCore case [13]) with less random noise, and this is represented in the model by a higher value of $A_C$ and fewer random links. The number of edges in the resulting graph is generally very close to that in the dataset that is being modeled; for instance, in the case shown in Fig. 3A the number of edges is 2% larger than the number in the Ito dataset (the results are similar for ItoCore and Uetz, see the Supporting Information). Although this represents only one method of adding random noise to the system, other random linking strategies (such as adding a random link to every node in the graph regardless of connectivity) yield similar results (see Supporting Information). In this model, the exposure of two different surfaces for individual proteins in the Uetz and Ito experiments represents creating graphs from 2 independent realizations of the MpK model, holding $p$ fixed for each model protein but resampling $K$ independently. Two model graphs sampled in this way exhibit a very low correlation between connectivities as expected (Fig. 3B). In this case the value of $R^2$ is 0.012, an order of magnitude smaller than that observed for Ito vs. Uetz datasets, indicating that sampling of the surfaces in the two experiments are most likely not completely unrelated to one another. Nonetheless the above results demonstrate that a purely physical model can produce networks that are unrelated topologically but nonetheless scale-free from a single population of proteins.

Recent studies have indicated that topological properties aside from the degree distribution also exhibit interesting scaling behaviors in the PPI and other biological networks [6,17,29-31]. Perhaps most interesting is the fact that the clustering coefficient of a node (a measure of the tendency of a node's neighbors to contact one another, denoted $C$) scales with the degree of the node; i.e. $C(k) \sim k^{-2}$ [17,29,31]. This finding has been explained in terms of a tendency for such networks to display "hierarchical modularity," but we find that our purely physical model displays similar scaling behavior in the absence of any considerations of modularity (see the Supporting Information). We also find that other, higher-order features of the graph, such as the relationship between the connectivity of a node and the average connectivity of its neighbors [30], is also observed in our physical model (see the Supporting Information). It is thus clear that interpretation of global topological features in light of evolutionary or functional pressures is difficult to evaluate in the absence of purely physical, non-evolutionary controls, and these results highlight the potential utility of our model as a "null model" for understanding such observations in the future. Although it has also been shown that more local properties of a graph may potentially contain interesting evolutionary traces [32,33], we leave exploration of those properties to future work.

**Correlation between Connectivity and Hydrophobicity**

Although the above graph theoretic results are suggestive, our model makes another key testable prediction that explicitly relates the MpK model to the physical reality of protein-protein binding. Specifically, our model suggests that a relationship should exist between the connectivity of a protein in the PPI network and the surface hydrophobicity of that protein. In the case of the experimental data, we do not know which specific surface is exposed in the experiment; we only know the (approximate) value of $p$ for a subset of proteins. In the MpK model, it is clear that the average value of $K$ will follow

$$\langle K \rangle = Mp, \tag{2}$$

with a standard deviation ($\sigma_K$) of

$$\sigma_K = \sqrt{M(p)(1-p)}. \tag{3}$$

These features of the binomial distribution of $K$ indicate that averaging over populations of proteins with similar values of $p$ should provide a method for overcoming the inherent uncertainty in the relationship between $p$ and $K$ (especially at values of $p$ around 0.5, where $\sigma_k$ is maximal). We thus expect a strong correlation between <log($k$)> and <$p$> at some bin size in $p$ and a weak correlation between $\log(k)$ and $p$ for individual proteins (given that affinity, not free energy, determines connectivity, the $\log(k)$ gives a stronger correlation than $k$). The model also predicts that $\sigma_K$ will increase with increasing $p$ up to $p = 0.5$.

The above analysis introduces a new parameter into the system; namely, the bin size in $p$ over which the averaging occurs. In general, larger bin sizes result in larger correlations but weaker statistical significance given the fact that fewer points are used to calculate the correlation. If we take the largest bin size in $p$ that nonetheless results in a statistically significant correlation (P-value < 0.05), we find that the correlations between $<\log(k)>$ and $<p>$ are 0.84 for ItoCore, 0.79 for Uetz and 0.17 for Ito (P-values of 0.012, 0.025 and 0.014, repsectively). The bin size for both ItoCore and Uetz in this case is 0.05 units in $p$ and for Ito is 0.001 (although it is important to note that a correlation of 0.74 exists for Ito at a bin size of 0.05, but the P-value in this case is 0.052, just above the P-value cutoff for significance). The maximum correlation is displayed for ItoCore and the ItoCore model graph (the graph with the degree distribution shown in Fig. 3A) in figures 4A and B (the maximum for the ItoCore model is 0.89 and also occurs at a bin size of 0.05). The dependence of R on the bin size is similar between the model and the data (Fig. 4C for ItoCore and the Supporting Information for Uetz and Ito), although the correlation at intermediate bin sizes is somewhat larger in the model in all cases. This is likely due to the fact that $p$ is only approximately known for the proteins in the datasets but is exactly known for the model and the fact that every hydrophobic residue contributes equally to binding, whereas for the experimental PPI networks more bulky hydrophobic residues may contribute more to stickiness and thus to connectivity. In the case of the Ito data, our random linking model indicates that there is a significant amount of noise at low values of $k$ (especially for those nodes with $k = 1$).  Consistent with this finding, the maximal correlation between $<\log(k)>$ and $<p>$ in Ito increases to 0.89 (P-value 0.019) at a bin size of 0.05 when all $k = 1$ nodes are removed from the dataset.

It is important to note that the binning procedure does not produce statistically significant correlations between connectivity and other types of amino acids.  For instance, we observe no statistically significant correlation between the percentage of charged amino acids (DEKR) on the surface and connectivity, despite the fact that such residues have been implicated in the determination of specificity in protein-protein interactions.  The raw correlation between % of charged residues on the surface and the log of connectivity is –0.09 for the ItoCore data set, and we observe no statistically significant correlation at any bin size (see Fig. 4C, P-values in this case are between 0.15 and 0.3 for all bin sizes, indicating a lack of statistical significance).  As a further control we calculated the correlation between connectivity and the % of eight randomly chosen amino acids, and again find no statistically significant correlations at any bin size (data not shown).  From these results it is clear that binning alone does not guarantee strong and statistically significant correlations.

A clear and non-trivial prediction of our model is that the standard deviation in the log of the connectivity will increase as the hydrophobicity of the surface increases up to a maximum at 0.5.  This arises from the fact that higher values of $p$ simply represent the *possibility* that a protein will expose a large number of hydrophobic residues (and thus exhibit a large connectivity) but this does not ensure that the subset of residues that actually are involved in binding are actually hydrophobic.  Consistent with this prediction, we find an increase in the dispersion in connectivity with increasing $<p>$ (see Fig. 5B for the ItoCore and model results). This is true for all of the experimental data sets (data not shown). The strong correlations we observe between connectivity and hydrophobicity in the Y2H data and the close correspondence we find between the experimental and theoretical results lend strong weight to our physical picture.

## Discussion

The model discussed above represents an extremely attractive alternative to evolutionary models of these graphs given that this model is inherently simpler and is based on very basic physical properties and not elaborate evolutionary mechanisms. This model also explains a number of the observations that have been made regarding PPI networks: the existence of scale-free networks in very noisy experiments, the lack of correlation between degrees in Uetz and Ito, the promiscuity of baits when compared with preys and the scaling of $C$ with $k$. The model accomplishes this based on only one fittable parameter ($A_C$). To our knowledge no evolutionary model has exhibited all of the above features. Indeed, evolutionary models that produce pseudo-bipartite structures [34] are inherently unable to reproduce the observed $p(C)$ distribution and $C(k)$ behavior since each node in these networks has C=0. Finally, the correlation of $k$ with the fraction of hydrophobic surface residues (and the strong similarity in the behavior of this correlation between the model and the data as a function of the bin size in $p$) is a straightforward demonstration of the feasibility of our physical model.

Although the results of our model are very suggestive, our findings do not imply that the PPI experiments or especially curated online PPI databases do not contain any relevant biological or evolutionary information at all. Indeed it is possible to find weak correlations between biological observables and PPI network quantities [24] just as it is possible to find cases in which network features replicate findings from more careful biological studies. These weak correlations are completely consistent with a picture in which the majority of links in the network are the result of non-specific interactions or experimental noise. Our results strongly indicate, however, that the interpretation of graph theoretic features of high-throughput experiments in the light of evolutionary processes must be tempered by the exploration of alternative physical hypotheses; indeed, this physical picture represents a null model against which future results regarding PPI networks should be measured. The model discussed above, while very important in terms of the PPI network, might also be employed in various forms to describe other types of scale-free networks. Physical models based on additive or multiplicative processes could be used to provide explanations for many scale-free graphs, especially those that involve networks of macromolecules that bind one another.

**Acknowledgements**

The authors would like to thank Dr. Nikolay Dokholyan, Dr. Boris Shakhnovich and Dr. Leonid Mirny for their comments on the manuscript. EJD acknowledges support from a Howard Hughes Medical Institute predoctoral fellowship and the authors acknowledge support from the National Institutes of Health.

Correspondence and requests for materials should be addressed to eugene@belok.harvard.edu

**Figure Legends**

**Fig. 1.**  Correlation Between PPI Networks  **A.**  The correlation between the network degree of a given protein in the Ito and Uetz datasets.  Each point corresponds to a particular protein that exhibited interactions in both experiments.  **B.**  A plot similar to **A** but comparing the ItoCore dataset with Uetz.

**Fig. 2.**  A Physical Model for Protein-Protein Interaction Measurements  **A.**  A schematic of the model described in the text.  Association free energies are largely the result of desolvation of the two protein surfaces.  The overall burial of hydrophobic groups is represented by the sum of the contributions from each protein  **B.**  The distribution of surface hydrophobicities in Yeast proteins.  The fraction of surface residues that are hydrophobic (defined as residues AVILMFYW) is calculated according to the description in the Methods section of the Supporting Information.  This distribution is taken from proteins in the Ito experiment.  The red squares represent the model hydrophobicities sampled from a Gaussian distribution with the same mean and standard deviation as the Ito proteins themselves.  **C.**  A degree distribution for the realization of the model used in **B** and **C**.  The cutoff was chosen such that the power-law fit gives an exponent of around –2.0, close to that of Ito graph.  The degrees in this plot are shifted by +1 to allow for orphans (nodes of degree 0) to be displayed on a log-log plot.  Note that the fraction of orphans in the graph is very high.

**Fig. 3.**  Degree Distributions and correlations for Model PPI Networks  **A.**  Comparison of degree distributions for the Ito dataset and a realization of the random linking model. In this case all orphans from the model graph of 3200 nodes are connected to one node that does exhibit connections randomly.  The line represents a power-law with an exponent of –2. The degrees in this plot are not shifted as they are in Fig. 2C. **B.** The correlation between degrees for in the model of Ito compared to the model of Uetz. In this case the different experiments are represented as independent sampling of values of $K$ from a population of proteins with the distribution of $p$ values shown in Fig. 2B. The Ito model is equivalent to the random linking results in panel **A**, and the Uetz degrees are taken from its random linking model (the degree distribution for that graph may be found in the Supporting Information). The linear correlation is 0.04 in this case.

**Fig. 4.** Correlations between hydrophobicity and connectivity. **A.** The maximal correlation between <log($k$)> and <$p$> for the ItoCore dataset; the correlation is 0.84 with a P-value of 0.012. The bin size in $p$ in this case is 0.05. **B.** The maximal correlation between <log($k$)> and <$p$> for the model of ItoCore; the correlation is 0.87 with a P-value of 0.014 at a bin size of 0.05 units in $p$. **C.** The dependence of the correlation between <log($k$)> and <$p$> as a function of the bin size in $p$ used to define the populations over which $p$ and log($k$) are averaged. The dependence for hydrophobic residues is very similar between the ItoCore data and the physical model for ItoCore. In the case of the charged data set $p$ is taken to be the percentage of charged residues on the surface and the correlation dependence is calculate exactly as for the hydrophobic residues. None of the correlations for the charged dataset are statistically significant. **D.** The maximal correlation obtained for the Ito dataset when nodes with $k = 1$ (the nodes predicted to be most susceptible to random linking noise by our model) are removed. The correlation is 0.89 in this case with a P-value of 0.019.

**Fig. 5.** Standard Deviation in Connectivity. As predicted by the MpK model, in both the model and the ItoCore network the standard deviation of log($k$) in a given bin in $p$

increases with increasing $<p>$ for that bin (the bin size is set at 0.05 for both the model and experimental networks in this figure). The lack of a maximum at $p = 0.5$ (as predicted by equation 3 in the text) is due to the fact that very few proteins exist in those bins with large $<p>$, decreasing the standard deviation for the most hydrophobic bin in each case.

**Figure 1**

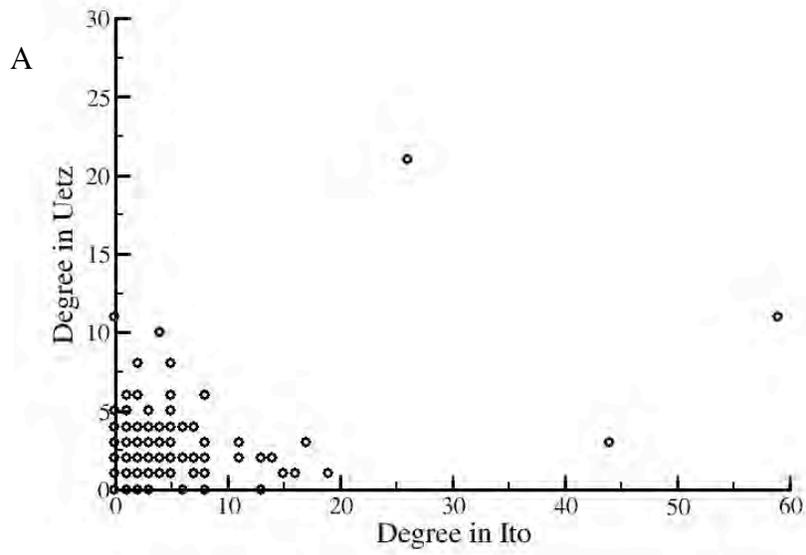

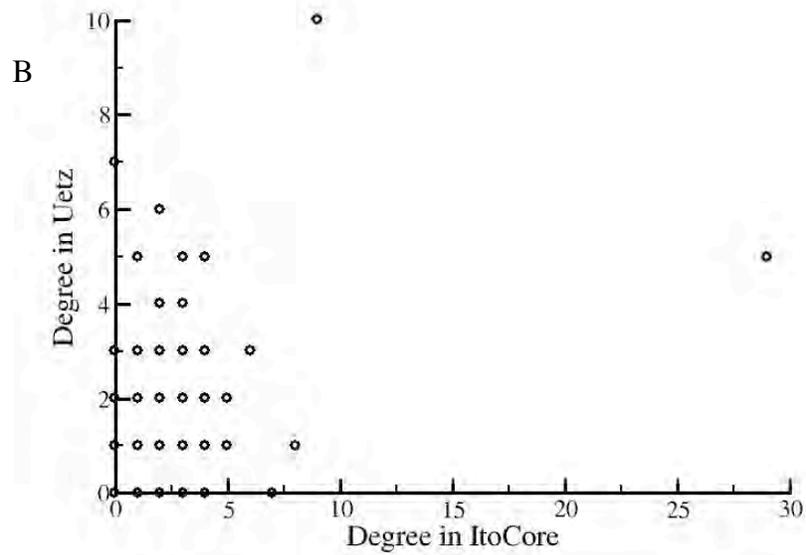

**Figure 2**

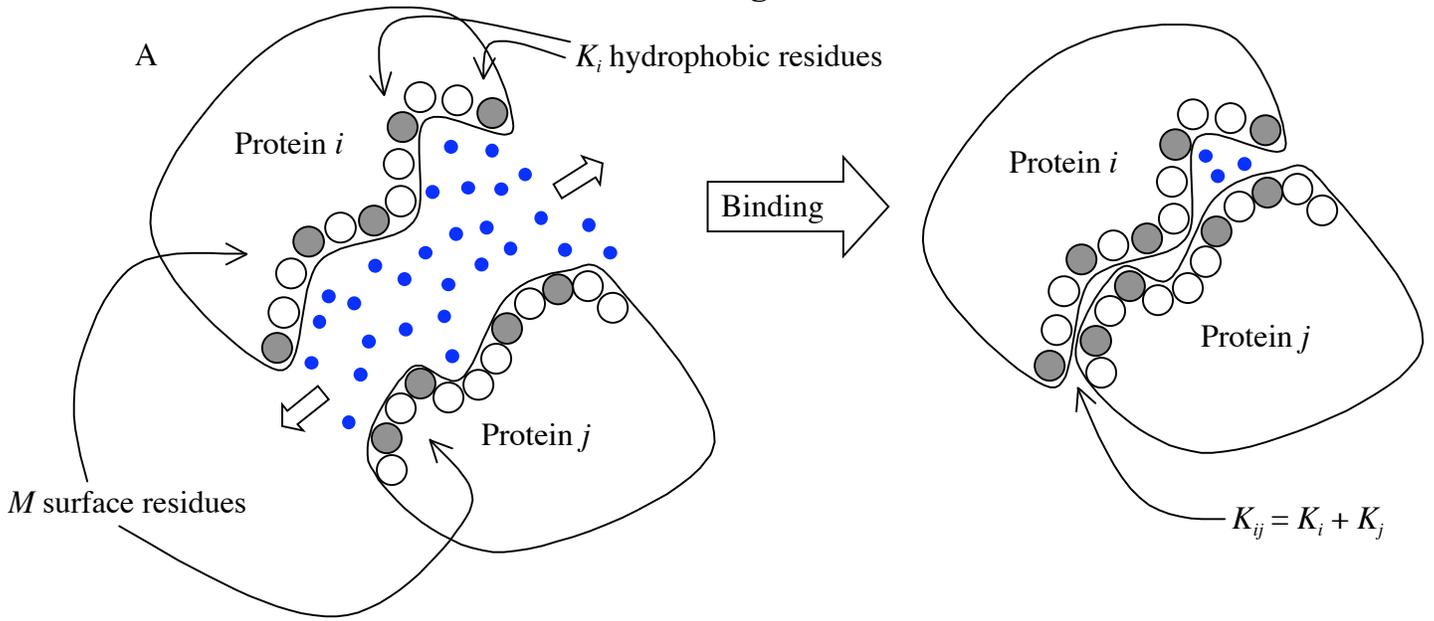

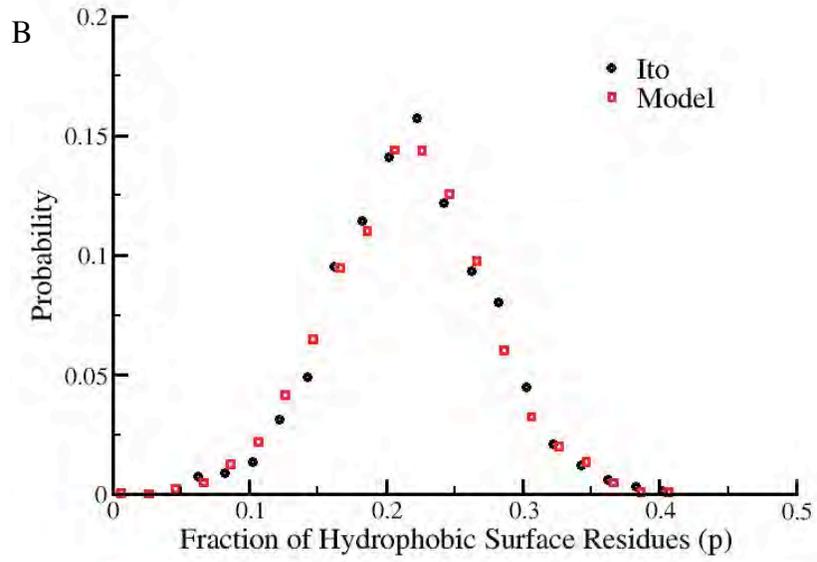

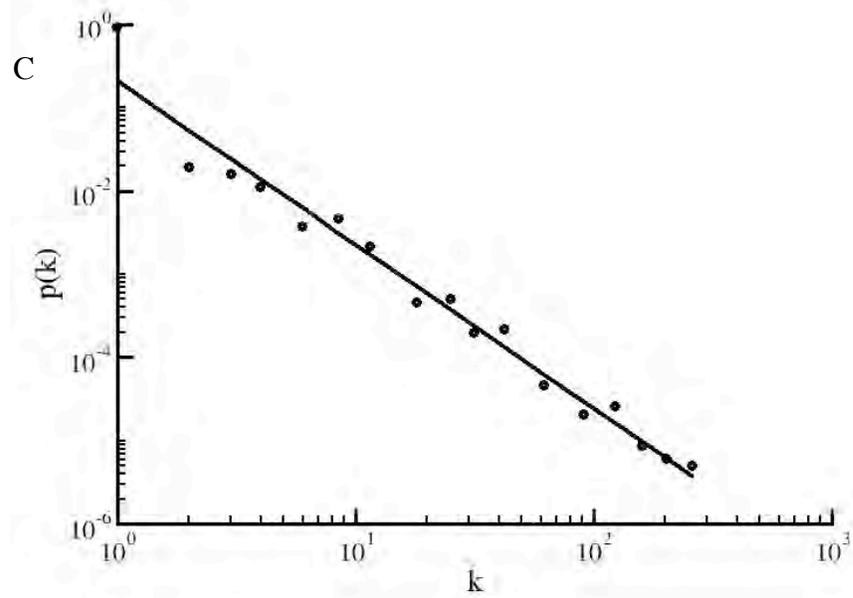

**Figure 3**

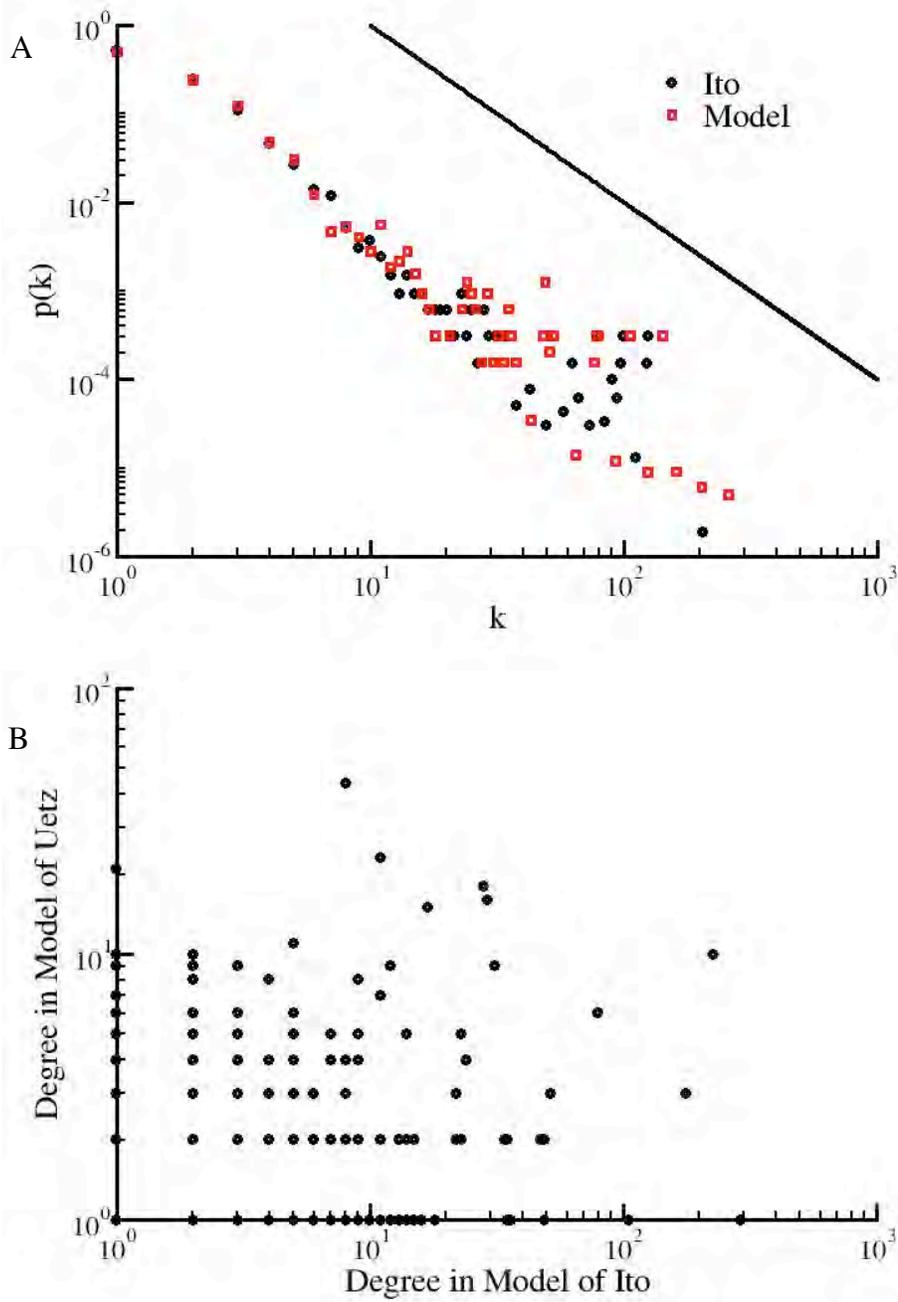

**Figure 4**

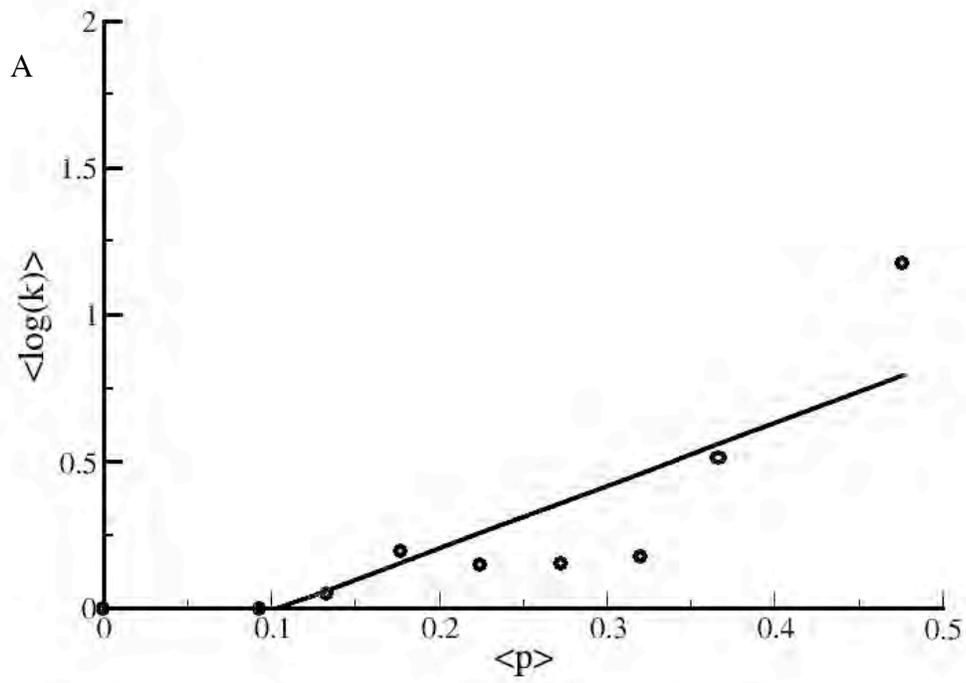

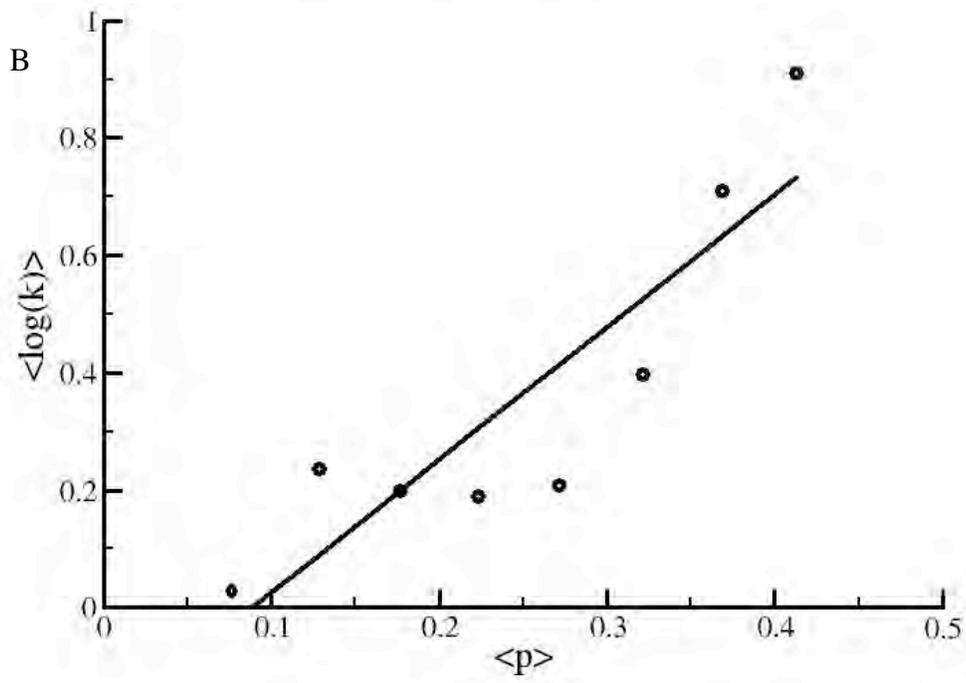

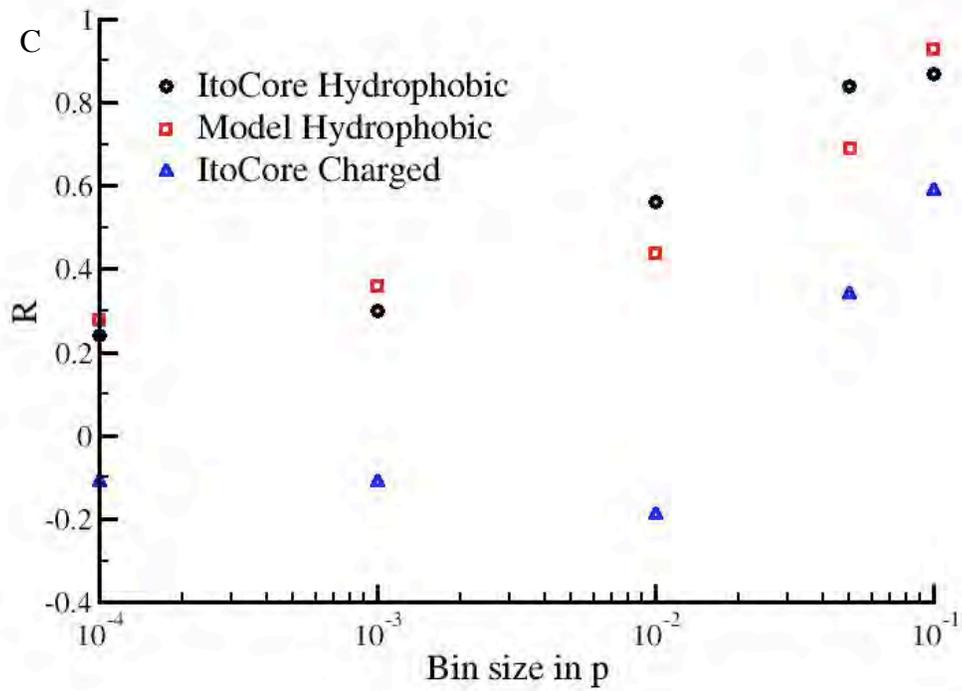

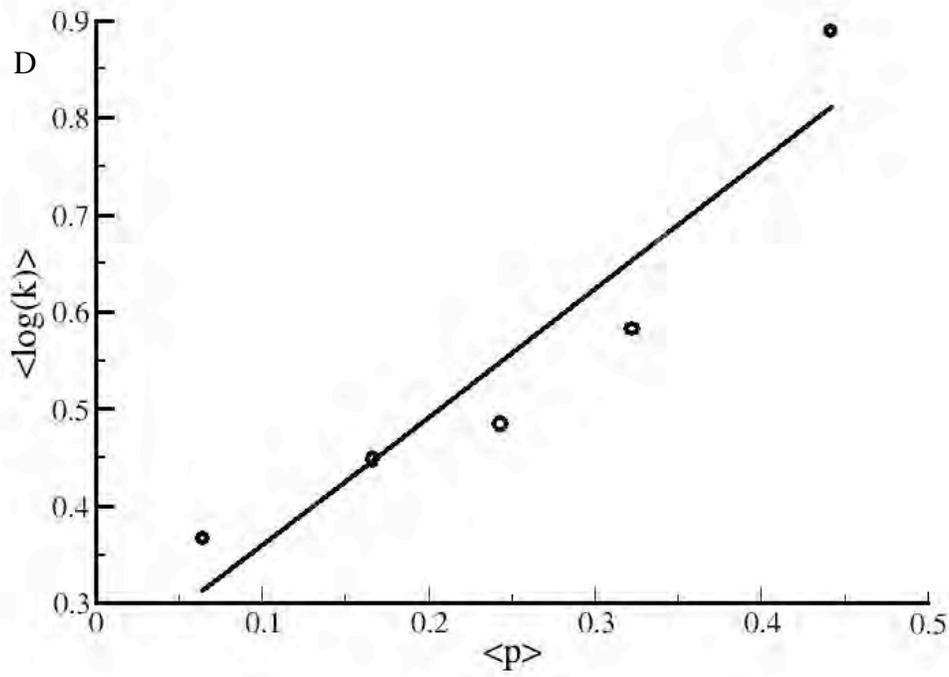

**Figure 5**

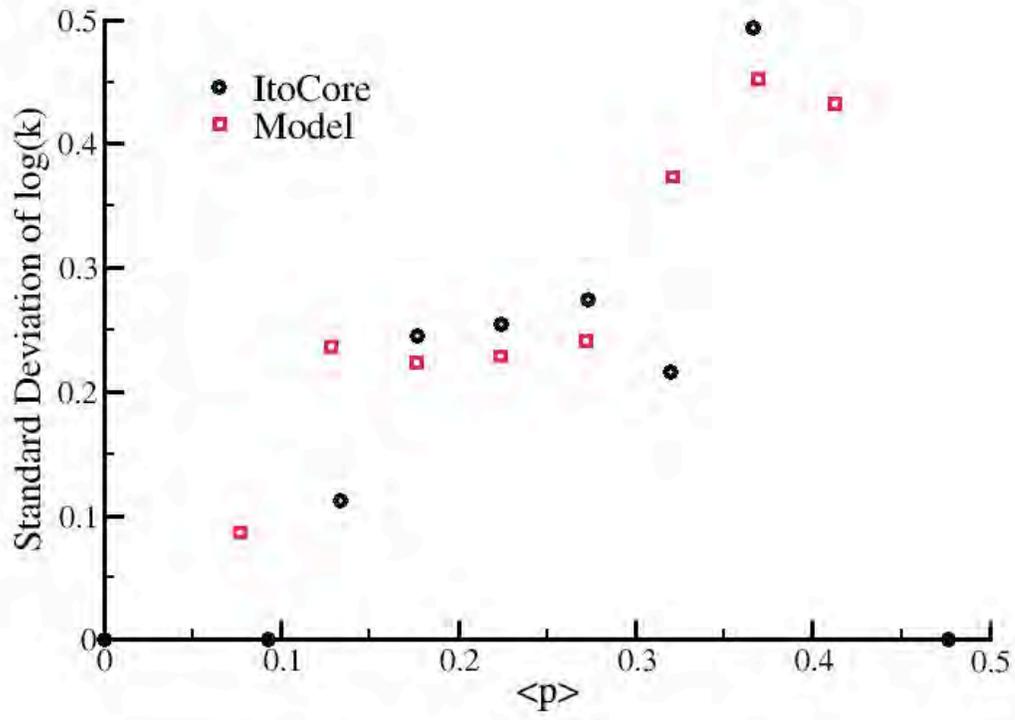

# Supporting Information

## Distribution of $\gamma$ for multiple realizations of the MpK model

To determine how representative the realization of the MpK model employed for Fig. 2C and Fig. 3 in the text is compared to a population of graphs produced under this model, we create 100 realizations of the MpK model. In this case we utilize a single realization of the values of $p$ for each model protein; the overall degree distribution results described below do not change when multiple realizations of the value of $p$ for each model protein are employed (data not shown). The 100 separate realizations used for this analysis each represent a re-calculation of $K$ for each protein according to the binomial process described in the text. The values of $M$ and $A_c$ are fixed for each realization using the exact parameters used for the realization shown in Fig. 2C in the text.

We find that each of these 100 realizations produces a graph with a degree distribution well-fit by a power law function (for the degree distributions of all 100 realizations see Fig. S1A). The distribution of the power-law exponents for all realizations is approximately Gaussian (see Fig. S1B) with an average of 2.05 and a standard deviation of 0.10. The realization displayed in Fig. 2C in the text is representative of the degree distributions in this ensemble of graphs (the value of $\gamma$ in that case is well within a single standard deviation in $\gamma$ for the ensemble).

## Relationship between $M$ and $A_c$

The model proposed in the text apparently has two parameters: $M$ and $A_c$. The value of $M$, however, does not effectively change the behavior of the model; it simply sets the value of $A_c$ that is required to obtain a scale free network with a particular degree distribution. To demonstrate this explicitly, we employ a single realization of $p$ values for a set of model proteins and calculate a single realization of $K$ for that set of model proteins with values of $M$ from 30 to 200. For each realization we build a graph at a $A_c$ such that the fitted $\gamma$ for the network is as close to 2.0 as possible; in this case the maximal deviation from 2.0 is 0.05. The value of $A_c$ required to obtain this behavior in the degree distribution is an exponential function of $M$ with $A_c = 33.29*\exp(0.74*M)$ (see Fig. S2). The correlation coefficient of the exponential regression is 0.998, indicating a nearly perfect relationship between $M$ and $A_c$. This finding indicates that, for any desired behavior of the network, the value of $M$ specifies the value of $A_c$ required to obtain that behavior (as long as $M$ is sufficiently large and the desired network behavior is achievable within the confines of the model). Given that the dependence of $\gamma$ on $A_c$ is monotonic in nature (see below), it is clear that $M$ exactly specifies the value of $A_c$ required to obtain a particular exponent. Thus, despite the existence of two parameters in the model, these parameters are not independent of one another, resulting in a single effective parameter.

**Dependence of $\gamma$ on $A_c$**

As discussed above, the MpK has effectively a single fitable parameter, and if we fix $M$ (i.e. $M = 100$ as in the text), this single parameter is the buried hydrophobic surface area cutoff ($A_c$) at which the graph is constructed. To test the dependence of our results on this cutoff we calculate graphs at a range of cutoffs. All graphs in this region are well fit by power-law functions, although the quality of the fits decays as the cutoff becomes larger and the total number of points decreases. The dependence of the exponent on the cutoff is shown in Fig. S3.The relationship between the cutoff and the exponent is monotonic, with $\gamma$ increasing as the stringency of the cutoff increases. Graphs with power-law exponents between 2 and 2.7, i.e. similar to those observed in the Ito, ItoCore and Uetz data sets (and many other scale-free networks) [1-3], are observed over a range of cutoff values.

**Degree distributions of ItoCore and Uetz model**

Models for both ItoCore and Uetz are constructed according to the algorithm described in the text. In the ItoCore case, the values of $p$ and $K$ are identical to the values used for Ito, consistent with the fact that ItoCore represents a subset of the Ito experiment. For the Uetz model, the same values of $p$ are used, but a different realization of $K$ values is employed. In each case the value of $A_c$ is determined by fitting the $\gamma$ of the resulting graph to the value of $\gamma$ observed in the corresponding experimental graph. The higher value of $A_c$ required to create ItoCore from the Ito dataset is expected from the nature of the ItoCore data; the ItoCore dataset is based on those interactions that exhibited a larger number of ISTs in the experiments [2], which quite naturally corresponds to a higher cutoff (i.e. a higher affinity is required to observe three IST tags as opposed to one). A number of random links were added to orphans such that the number of connected nodes in each graph matched the corresponding experiment; the resulting number of edges in each graph is similar to the experimental result. The comparison of the degree distributions is shown in figures S4A and S4B for ItoCore and Uetz, respectively. The degree distribution in Fig. S4B corresponds to the realization employed for the correlation between the Ito and Uetz models in Fig. 3B in the text.

**Alternative Random Linking Model**

To test the susceptibility of our model to other forms of random linking, we create an additional model of the Ito experiment in which a single random link is added to every single node in the graph regardless of connectivity. In this case every node is randomly linked to another node that is chosen with equal probability from all other nodes in the graph. The addition of a single "noisy" link in this manner produces graphs with very similar degree distributions to those observed in the original random linking model (see Fig. S5). The other features of the graphs produced by this model are also similar (i.e. p(C), C(k) scaling, etc., data not shown). Other implementations of random linking models (such as models in which a random link is added to each node in the graph but the random linking probability depends on the hydrophobic surface area of the "acceptor" node) yield similar results (data not shown).

**Clustering Coefficient Results**

Although the model presented in the text and above exhibits strong agreement with the experimental degree distributions, it is not clear that the above model will reproduce other well-known graph theoretic properties of the PPI system. One such feature is distribution of clustering coefficient in the graph [3]. The clustering coefficient of a given node $i$, or $C_i$, is defined in the following way:

$$C_i = \frac{k_{nn}}{\frac{k_i(k_i-1)}{2}},$$ **(S1)**

where $k_{nn}$ is the number of edges that exist between the neighbors of node $i$ and $k_i$ is the degree of node $i$. It is known that the $C$ of a node scales with the degree of the node, giving a pronounced $C(k)$ behavior that has been attributed to the evolution of hierarchical modularity in this and other networks[4-7]. The probability distribution of finding a node with a particular clustering coefficient ($p(C)$) for Ito and model Ito graphs are displayed in figure S6. The distributions are all similar to one another and to the results from random rewiring [8] of the respective graphs (data not shown).

As can be seen from figures S7, S8 and S9, the model discussed above reproduces the $C(k)$ scaling behavior of the experimental graphs. Randomly rewired [8] PPI graphs exhibit similar $p(C)$ behavior but do not exhibit the same extent of $C(k)$ scaling (data not shown). In the case of our physical interaction model the observed dependence of clustering coefficient on connectivity results from the fact that high connectivity nodes are inherently "sticky." The bulk of contacts made by such nodes are with nodes that have much lower inherent affinities, which basically precludes those nodes from contacting one another. Thus nodes of high connectivity will have low clustering coefficients, and the reverse will tend to be true for nodes of low connectivity. The $C(k)$ scaling observed in this system may thus simply be the result of physical interaction considerations and not from an evolved hierarchical or modular tendency in these graphs.

Another feature of the PPI graph that has been explained evolutionarily is the tendency for low-connectivity nodes to contact high-connectivity nodes and the tendency for high connectivity nodes not to touch one another [8-10]. The first observation may be obtained from our model on the basis of the arguments outlined above for $C(k)$. The second observation could also be obtained by implementing a "maximum" affinity cutoff due to the fact that over-expressing pairs of proteins that are inherently "sticky" may expose too much hydrophobic surface in the cell and cause aggregation. Indeed, our model recovers the scaling of average neighbor connectivity with the connectivity of a node [8] (see Fig. S10 for the Ito and model Ito results), indicating that our current model is sufficient to reproduce "higher-order" topological properties in the PPI network. The ItoCore and Uetz experimental and model results are similar to those shown for the Ito and Ito model (data not shown). Such arguments may explain the "compartmentalization" of the PPI network, but we leave more detailed analysis of this property to future work. It is important to note that the scaling in all of these properties of the PPI network is recovered from a model with a single tunable parameter that is fit only to the value of $\gamma$ and not changed in order to obtain correspondence in other properties.

**Correlation between Connectivity and Exposed Hydrophobicity**

As described in the Methods section below, we calculated the fraction of exposed residues that are hydrophobic in the Yeast proteins employed for the Ito and Uetz Y2H experiments [1,2]. In the model, we find that the $K$ of each node is correlated with the logarithm of its connectivity rather than with connectivity itself (data not shown). This empirical result can be explained in light of the fact that buried hydrophobic surface area determines the free energy of an interaction, but the observations in the experiment are a function of affinity (the exponential of free energy) rather than energy itself.

The comparison of correlation results between the model and the data for ItoCore is displayed in Fig. 4 in the text. The dependence of $R$ on bin size for the unmodified Ito and the Ito model is shown in Fig. S11A; the $k = 1$ deletion results for both Ito and the model are shown in Fig. S11B. The results for Uetz are shown in Fig. S12. The relationship between bin size and R is the same in all cases—as the bin size decreases, the correlation coefficient decreases and the P-value of the correlation (in general) also decreases. The correlation is somewhat lower at small bin sizes in the Uetz and ItoCore models when compared to the Ito model. In this case the difference is most likely due to the higher cutoffs employed to create the Uetz and ItoCore model graphs. These higher cutoffs remove a large amount of connectivity information from the graphs, which results in lower correlations at small bin sizes.  As discussed in the text, the MpK model predicts that the standard deviation in the $\log(k)$ will increase with increasing $<p>$ for some bin in $p$.  Consistent with this prediction we observe a very similar dependence of $\sigma_{\log(k)}$ on $<p>$ in both the ItoCore data and the ItoCore model (see Fig. 5B in the text).  The standard deviation results for the ItoCore and Uetz data and models are similar (data not shown).

**Methods**

In order to test the validity of our assumption that protein surfaces exhibit a Gaussian distribution of stickiness we calculated the surface hydrophobicities for the Yeast proteins employed in the Uetz and Ito experiments[12,13].  We obtained the sequences for proteins in these experiments from the SWISSPROT database[11].  We compared these sequences to sequences from proteins of known structure in the HSSP[12] database using BLAST[13].  Hits were considered significant if they exhibited E-values of less than $10^{-6}$. Hits were recovered for about 30% of the proteins in each experiment (679 for Ito, 182 for ItoCore and 246 for Uetz).  For those proteins with significant hits we determine the solvent accessibility of residues in the Yeast sequence using the solvent accessibilities of aligned positions in the HSSP sequence[12].  For the purposes of the calculations in this work we assume a residue is "exposed" if its solvent accessibility score[12] is greater than 50 (corresponding to contact with roughly 5 water molecules)[12,14], although changing the value of this cutoff from 30 to 100 does not change our distribution results significantly (data not shown).  The hydrophobicity of a surface is calculated as the fraction of all exposed residues (i.e. all residues with accessibilities greater than 50) that belong to the group of amino acids AVILMFYW.

**Figure Legends**

**Fig. S1.**  Ensemble of MpK graphs.  **A.**  Degree distributions for an ensemble of 100 graphs calculated where $K$ is independently determined in each case from a single realization of $p$ values for each model protein.  In this case $M$ is set to 100 and $A_c$ is held fixed.  The distributions are all well-fit by power-law functions.  **B.**  Probability distribution for the fitted exponents $\gamma$ of each degree distribution included in **A**.  The distribution is approximately Gaussian with an average of 2.05 and a standard deviation of 0.10.

**Fig. S2.**  Dependence of $A_c$ on $M$.  A single realization of the MpK model was calculated for a population of model proteins with fixed values of $p$ but varying values of $M$.  In each case $A_c$ was varied in order to obtain a graph with a fitted power-law exponent as close as possible to 2.0, and overall the graphs exhibit exponents within 0.05 $\gamma$ units of 2.0.  There is a strong exponential relationship between $M$ and $A_c$: the exponential fit of the data is indicated as a straight line in the log-linear scale of the figure.

**Fig. S3.**  Dependence of $\gamma$ on $A_c$.  To demonstrate the relationship between $\gamma$ and $A_c$ we calculated the power-law exponent of a single realization of the MpK model with $M$ set to 100 at varying values of $A_c$.  In this case exponents similar to those observed in the various Y2H PPI networks (i.e. $\gamma$ from 2 to 2.7) are observed over a range of $A_c$ values.

**Fig. S4**  Degree distribution results for models of the ItoCore and Uetz data sets.  **A.**  The degree distribution for the ItoCore graph compared to the model ItoCore graph constructed as described in the text.  The straight line in the graph represents a perfect power law with an exponent equal to 2.6 (taken from the fit of the ItoCore data).  **B.** The degree distribution for the Uetz graph compared to the model Uetz graph constructed as described in the text. The straight line in the graph represents a perfect power law with an exponent equal to 2.7 (taken from the fit of the Uetz data).

**Fig. S5** Additional random linking model. This figure represents the degree distribution of a random linking model in which a single "noisy," random link is added to every node in the graph. The degree distribution marked "Original RL Model" is taken from Fig. 3A in the text; the "Alternative RL Model" corresponds to the single noisy link model discussed here.

**Fig. S6.** The probability distribution of the clustering coefficient for the Ito and Ito model graphs (black and red lines, respectively).

**Fig. S7.** Scaling of the clustering coefficient with $k$ for the Ito graph and model. **A.** Scaling of $C$ with $k$ for the Ito dataset. The solid line in this panel and in panel C represents a power law with exponent –2 as previously reported. **B.** Scaling of $C$ with $k$ for the Ito model that gives the degree distribution in Fig. 3A of the text.

**Fig. S8.** Scaling of the clustering coefficient with $k$ for the ItoCore graph and model. **A.** Scaling of $C$ with $k$ for the ItoCore dataset. The solid line in this panel and in panel C represents a power law with exponent –2 as previously reported. **B.** Scaling of $C$ with $k$ for the ItoCore model that gives the degree distribution in Fig. S4A.

**Fig. S9.** Scaling of the clustering coefficient with $k$ for the Uetz graph and model. **A.** Scaling of $C$ with $k$ for the Uetz dataset. The solid line in this panel and in panel C represents a power law with exponent –2 as previously reported. **B.** Scaling of $C$ with $k$ for the Uetz model that gives the degree distribution in Fig. S4B.

**Fig. S10** Scaling of the average connectivity of a node's neighbors with the connectivity of the node itself. **A.** Results for the Ito graph. The straight line in the figure represents

a power law with an exponent of 0.6 as previously reported.  **B.**  Results for the random linking MpK model of the Ito data that gives the degree distribution in Fig. 3A in the text. The straight line in the figure is a power law with the same exponent as the line in **A**.

**Fig. S11**  Correlation results for Ito.  **A.**  The dependence of R on the bin size in $p$ for the raw Ito and model Ito graphs.  In the case of the Ito data, the correlations for bin sizes of 0.001 and lower are statistically significant (P-values < 0.05).  All of the correlations shown on the plot for the Ito model are statistically significant.  **B.**  The dependence of R on the bin size in $p$ for the modified Ito and model Ito graphs in which nodes with $k = 1$ have been removed.  Correlations for bin sizes of 0.05 and lower are statistically significant for the modified Ito data.  All correlations shown for the modified Ito model are statistically significant.

**Fig. S12** Correlation results for Uetz.  The dependence of R on the bin size in $p$ for the Uetz and model Uetz graphs.  In both cases correlations for bin sizes of 0.05 and lower are statistically significant.

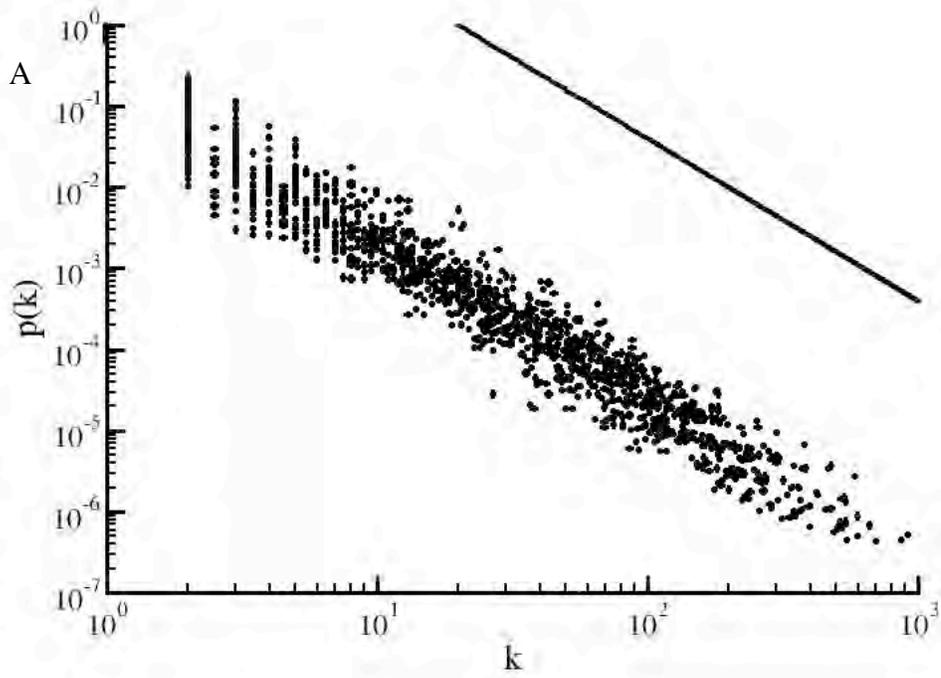

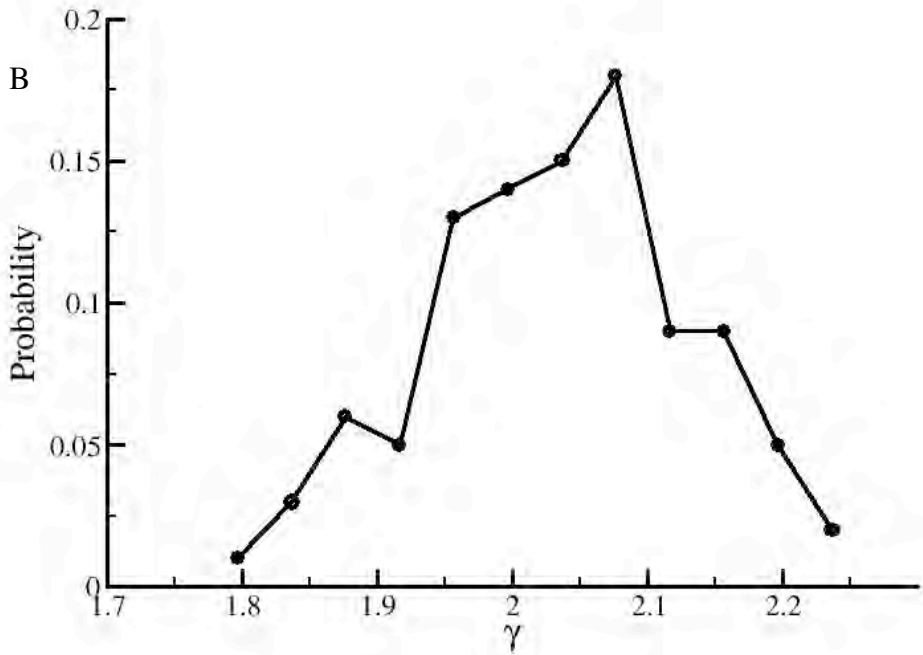

**Figure S2**

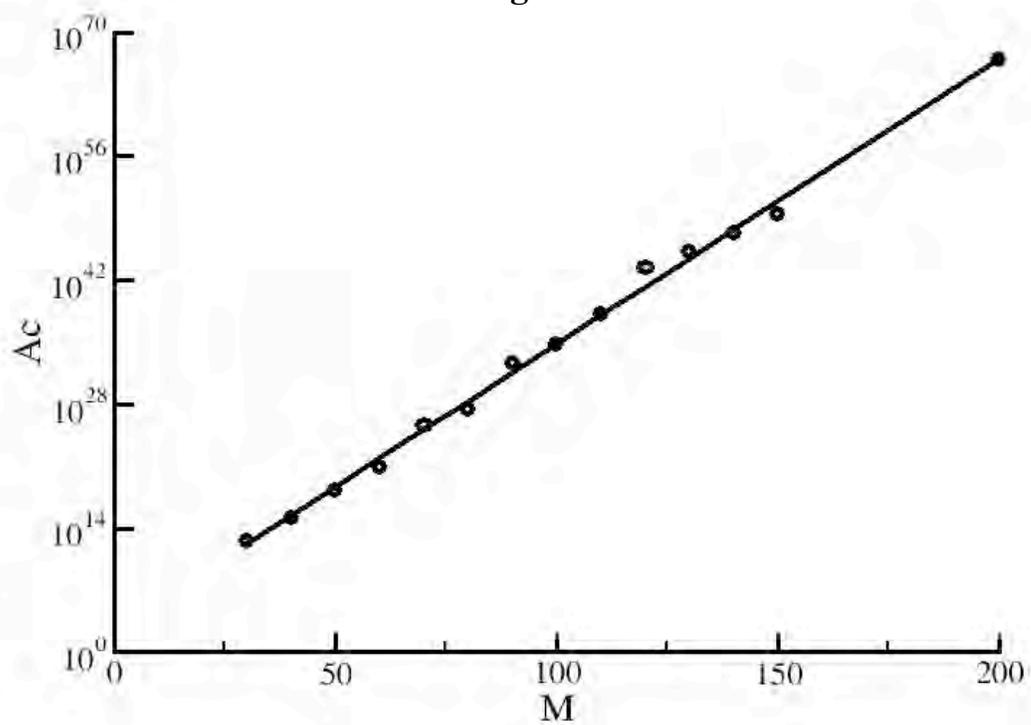

**Figure S3**

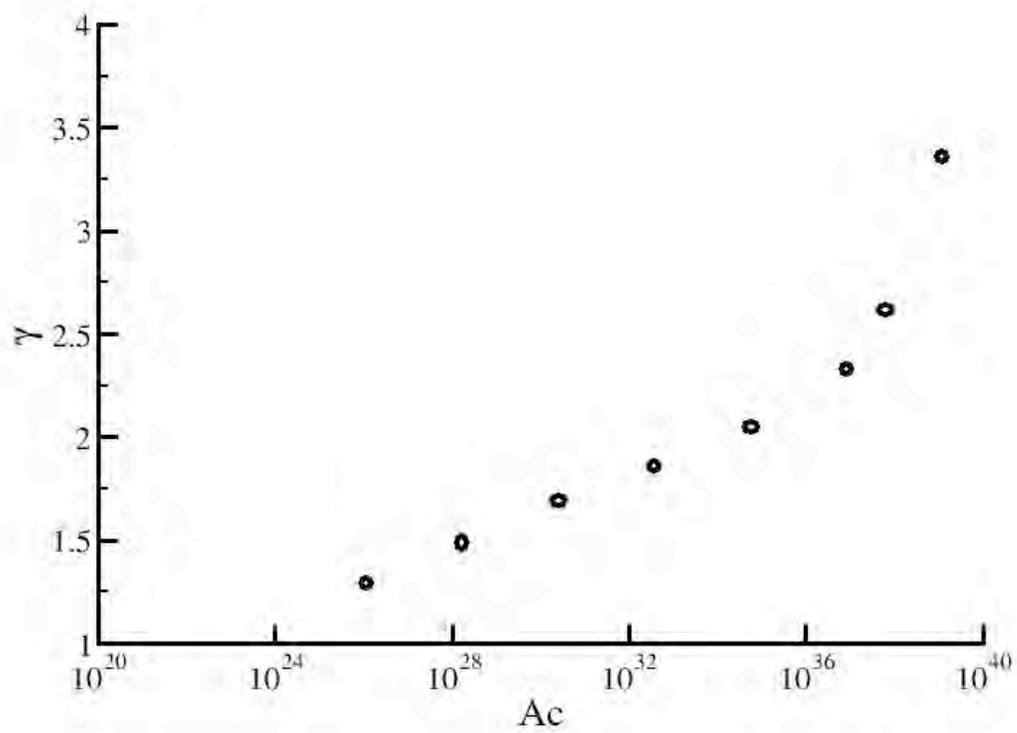

**Figure S4**

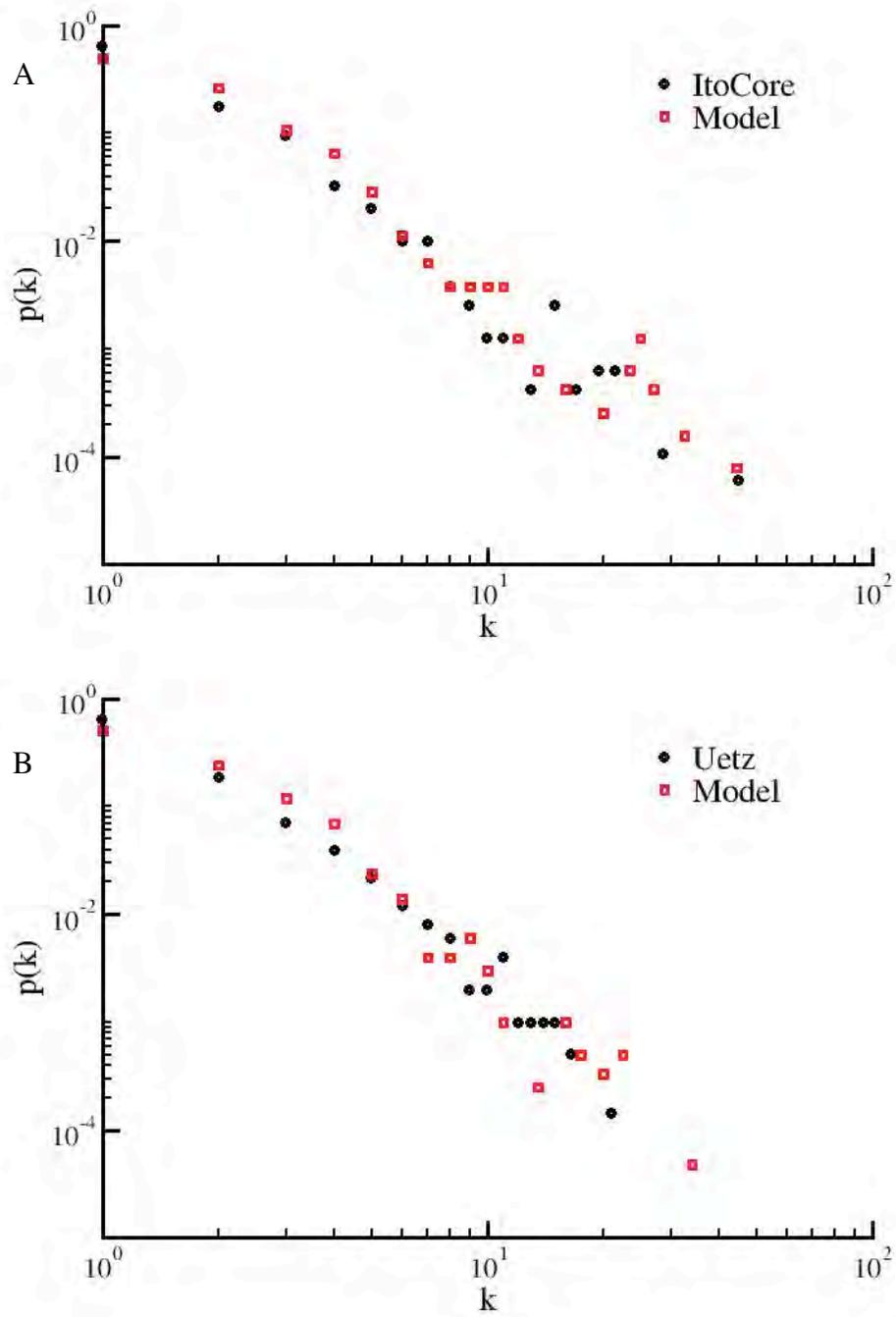

**Figure S5**

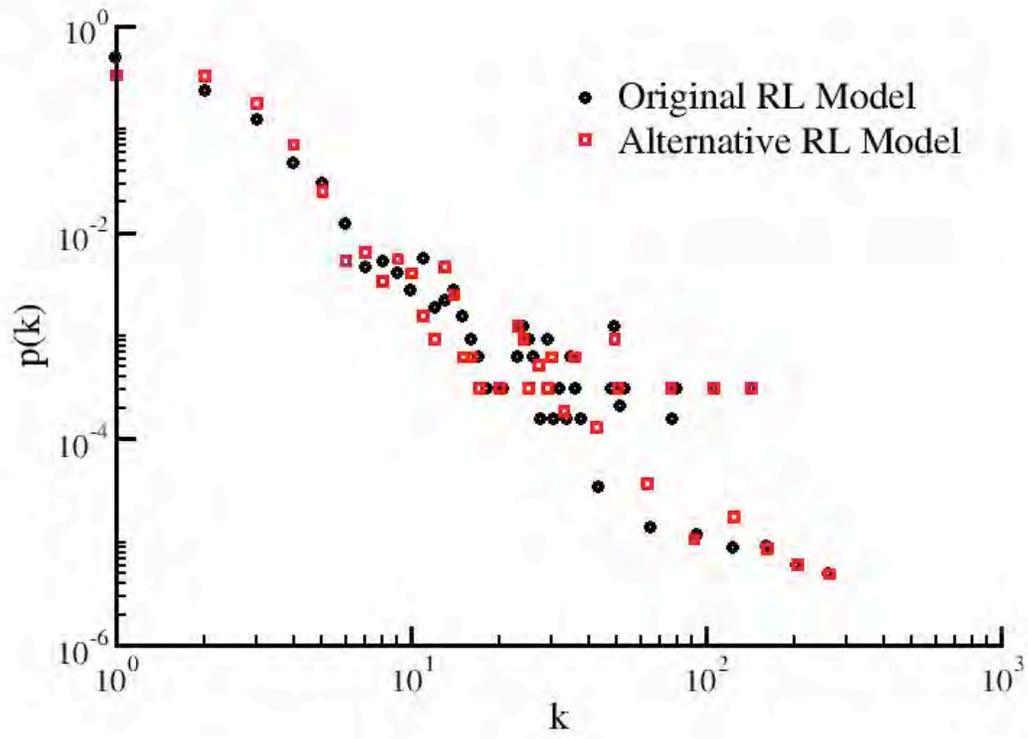

**Figure S6**

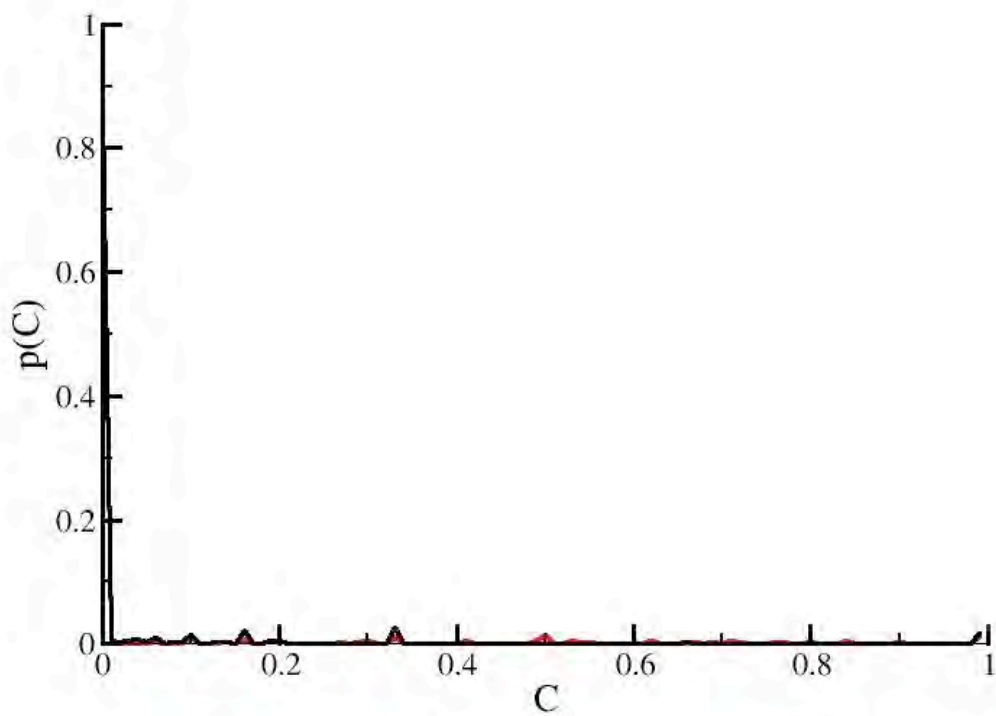

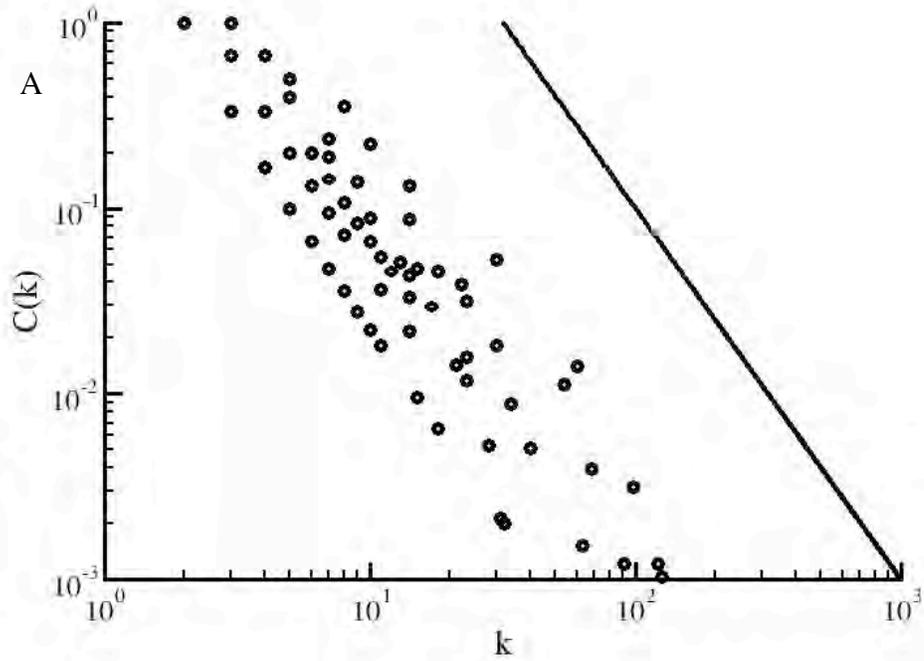

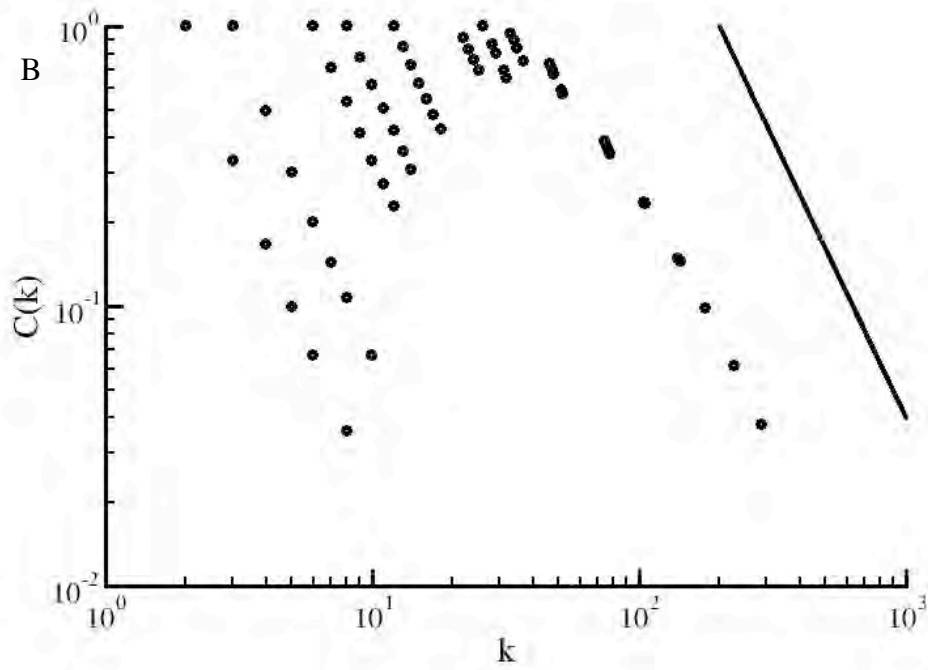

**Figure S8**

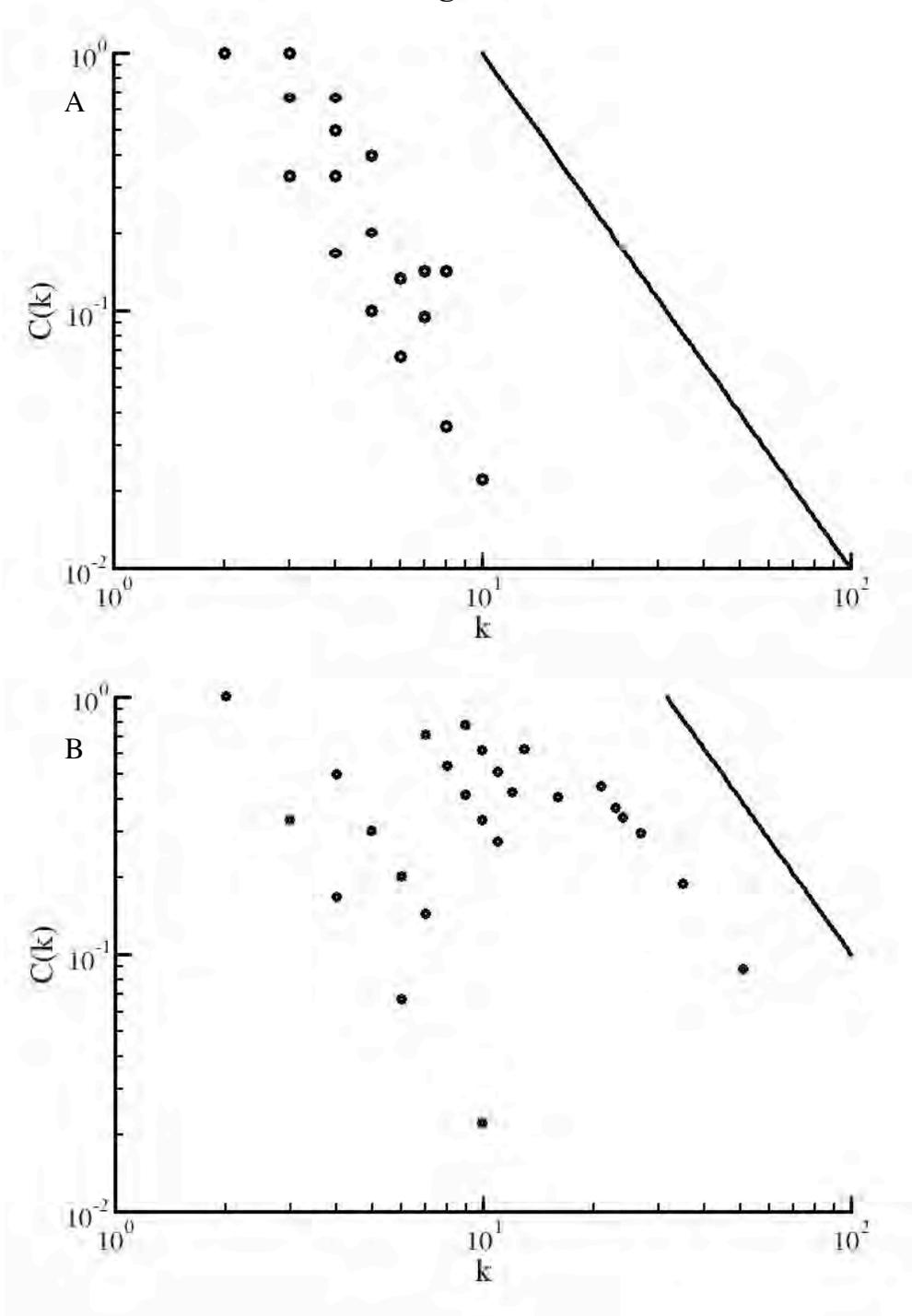

**Figure S9**

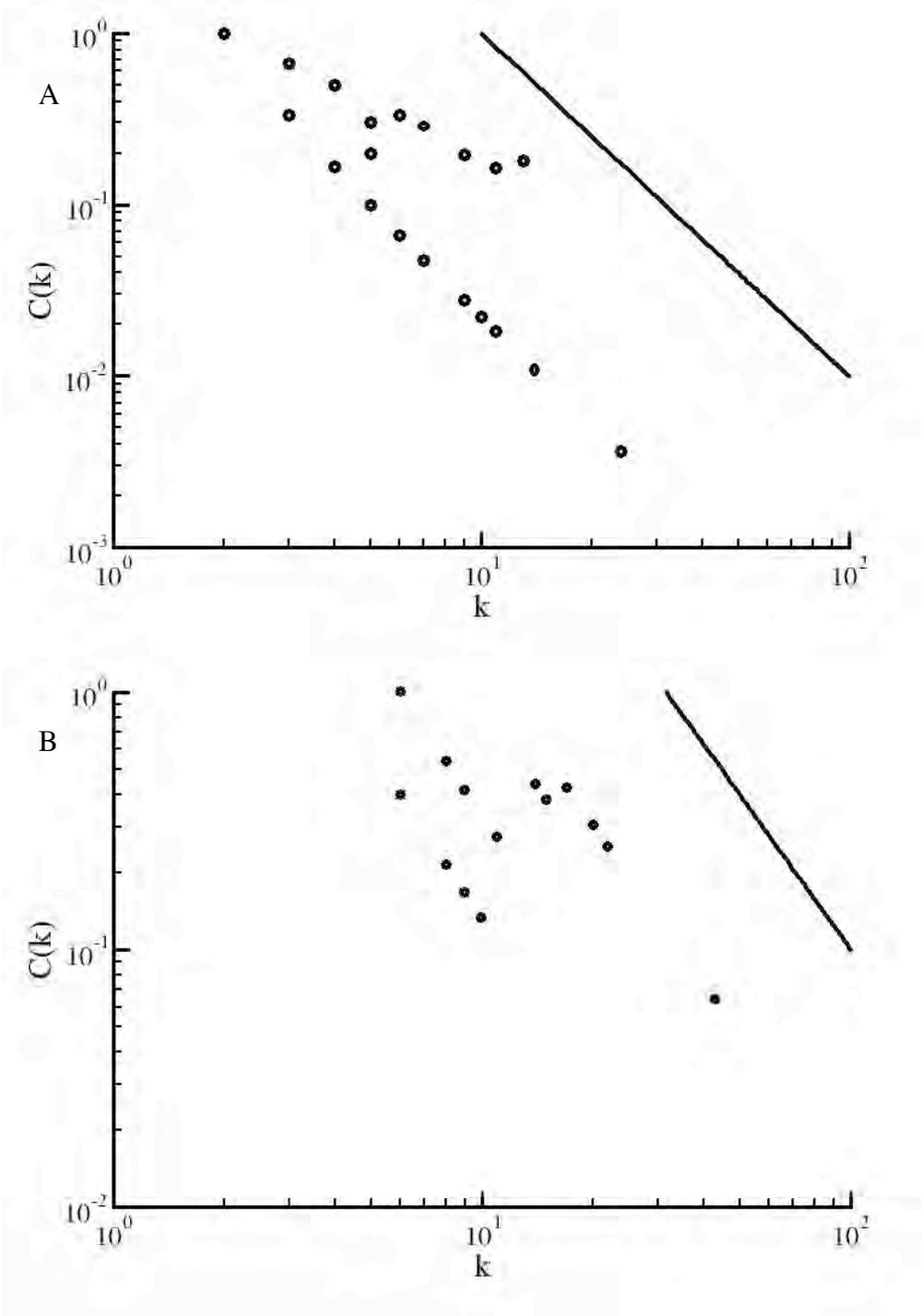

**Figure S10**

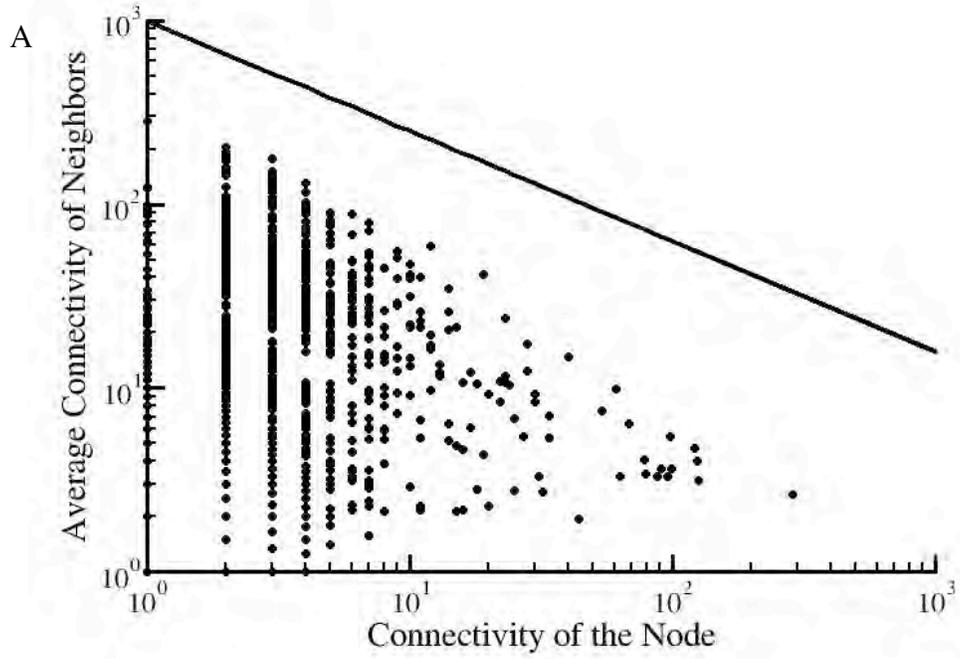

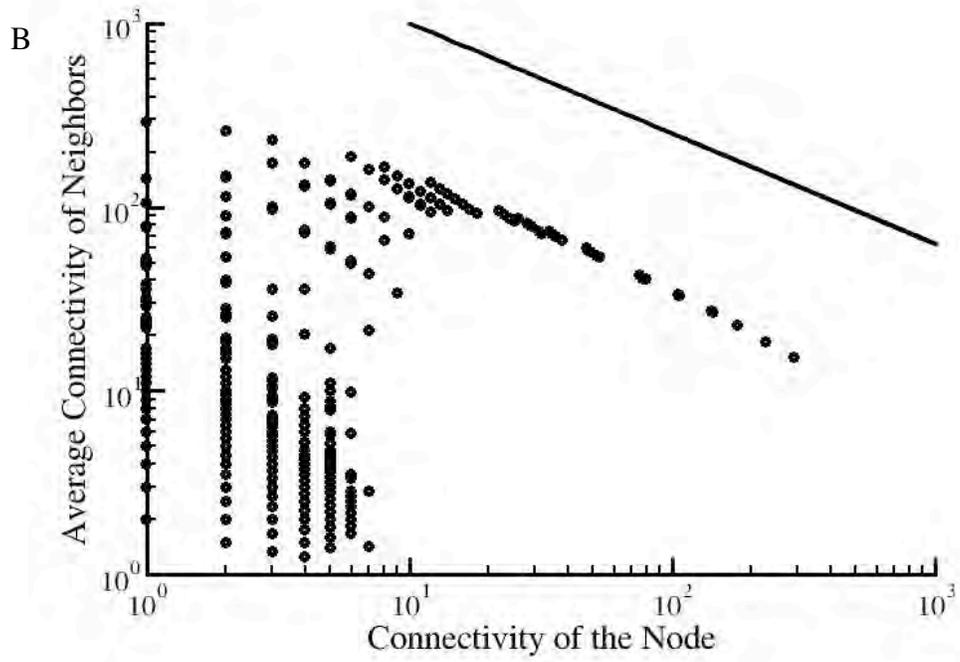

**Figure S11**

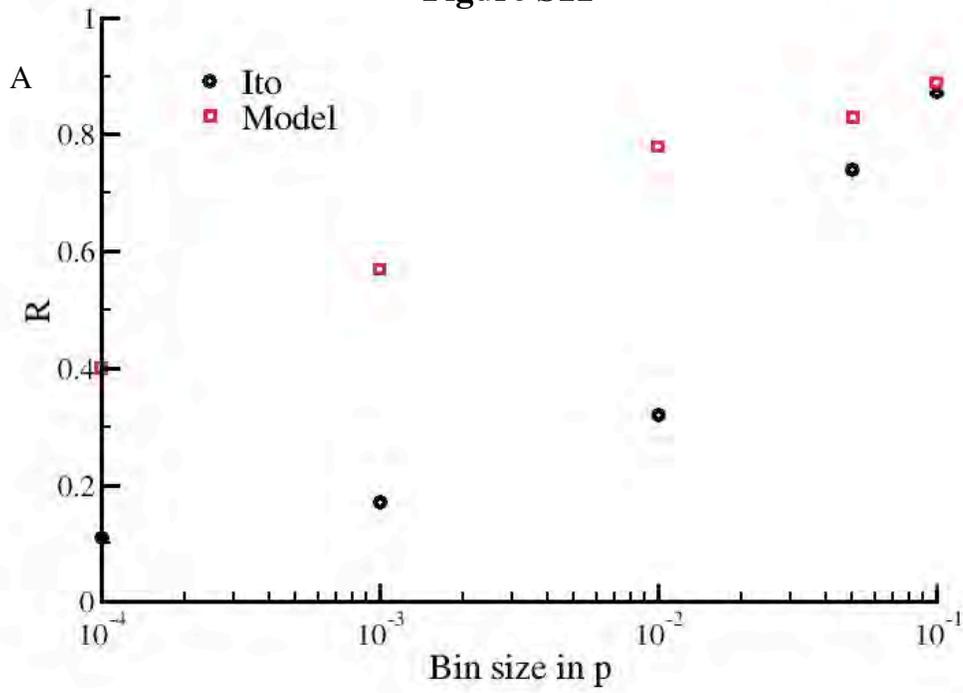

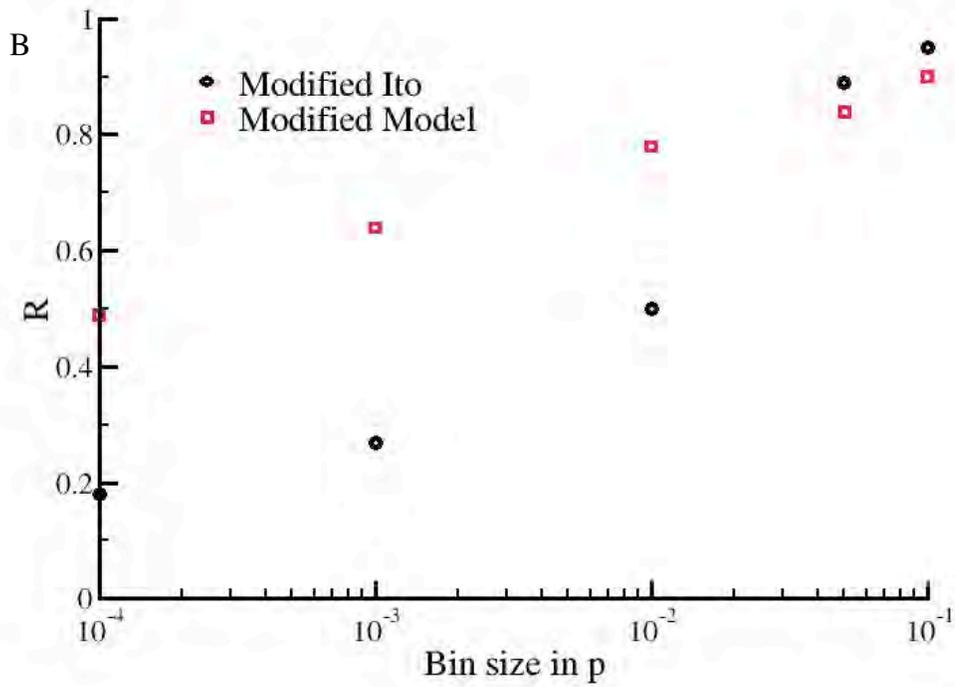

**Figure S12**

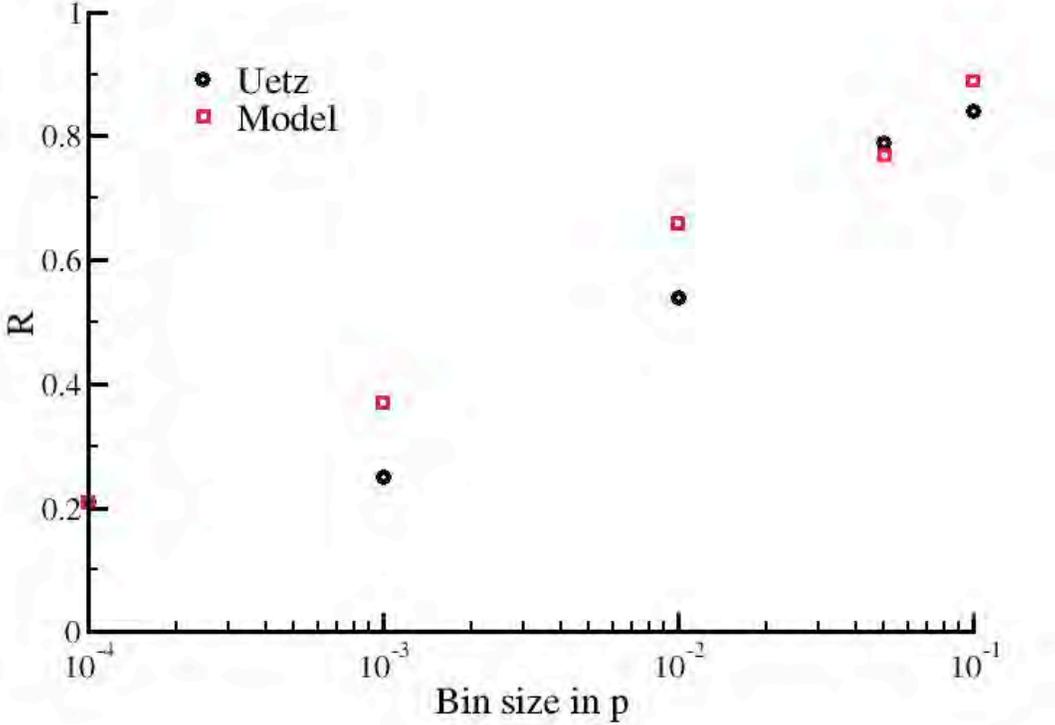